%% file: ms.tex
\documentclass[sigconf]{acmart}

\usepackage{booktabs} 
\usepackage[flushleft]{threeparttable}
\usepackage{xcolor}
\usepackage[utf8]{inputenc}
\usepackage{float}
\usepackage{amssymb}
\usepackage{ifthen}
\usepackage{multirow}
\usepackage{caption}
\usepackage{listings}
\usepackage{hyperref}
\usepackage{url,moreverb,xspace}
\usepackage{tcolorbox}
\usepackage{enumitem}
\usepackage{array,graphicx}
\usepackage{soul}
\usepackage{balance}
\usepackage{pifont}
\usepackage{etoolbox,siunitx}
\usepackage{nicefrac}
\usepackage{mathtools}

\usepackage{tikz}
\usepackage{pgfplots}
\usepgfplotslibrary{statistics}
\usetikzlibrary{fadings}
\usetikzlibrary{arrows.meta,shapes.geometric,shadows}

\usepackage{xcolor,pifont}
\newcommand*\colourcheck[1]{%
  \expandafter\newcommand\csname #1check\endcsname{\textcolor{#1}{\ding{52}}}%
}
\newcommand*\colourcross[1]{%
  \expandafter\newcommand\csname #1cross\endcsname{\textcolor{#1}{\ding{56}}}%
}

\usepackage[lined,boxruled,norelsize,linesnumbered]{algorithm2e}
\def\HiLi{\leavevmode\rlap{\hbox to \hsize{\color{gray!35}\leaders\hrule height .8\baselineskip depth .5ex\hfill}}}

\newtheorem{defn}{Definition}

\definecolor{lightgray}{gray}{0.85}
\definecolor{bananayellow}{rgb}{1.0, 0.88, 0.21}
\newcommand{\hlsmooth}[1]{{\sethlcolor{bananayellow}\hl{#1}}}
\newcommand*\rot{\rotatebox{90}}

\usepackage{graphicx}
\usepackage{subcaption}
\usepackage{seqsplit}

\newcommand{\tool}{\textsc{TEDD}\xspace} %
\newcommand{\baseline}{complete\xspace} %

\newcommand{\numberOfSubjects}{six\xspace} %
\newcommand{\totalNumberOfTests}{196\xspace} %
\newcommand{\praw}{PRAW\xspace} %

\SetKwComment{Comment}{$\triangleright$\ }{}
\SetCommentSty{itshape}
\newboolean{showcomments}
\setboolean{showcomments}{true}
\ifthenelse{\boolean{showcomments}}
{\newcommand{\nb}[2] {
  \fcolorbox{black}{gray!20}{\bfseries\sffamily\scriptsize#1:}
  {\sf\small$\blacktriangleright$\textit{#2}$\blacktriangleleft$}
}
}
{\newcommand{\nb}[2]{}
}

\newcommand{\head}[1]{\noindent\textbf{#1.}}

\newcommand{\code}[1]{{\texttt{#1}}}

\newcounter{fcounter}
\newcommand{\curl}[1]{\footnote{\url{#1}}}

\makeatletter
\newcommand{\thickhline}{%
    \noalign {\ifnum 0=`}\fi \hrule height 1pt
    \futurelet \reserved@a \@xhline
}
\makeatother

\renewcommand\footnotetextcopyrightpermission[1]{} 

\copyrightyear{2019} 
\acmYear{2019} 
\setcopyright{none}
\acmConference[Technical Report Series]{Technical Report}{2019}{FBK, Trento, Italy}

\begin{document}

\title{E2E Web Test Dependency Detection using NLP}

\author{Matteo Biagiola}
\affiliation{
  \institution{Fondazione Bruno Kessler}
  \city{Trento}
  \country{Italy}
  }
\email{biagiola@fbk.eu}

\author{Andrea Stocco}
\orcid{0000-0001-8956-3894}
\affiliation{
  \institution{Universit\`a della Svizzera Italiana}
  \city{Lugano}
  \country{Switzerland}
  }
\email{andrea.stocco@usi.ch}

\author{Ali Mesbah}
\affiliation{
  \institution{University of British Columbia}
  \streetaddress{2332 Main Mall}
  \city{Vancouver, BC}
  \country{Canada}
  \postcode{V6T 1Z4}
}
\email{amesbah@ece.ubc.ca}

\author{Filippo Ricca}
\affiliation{
  \institution{Universit\`a degli Studi di Genova}
  \city{Genova}
  \country{Italy}
  }
\email{filippo.ricca@unige.it}

\author{Paolo Tonella}
\affiliation{
  \institution{Universit\`a della Svizzera Italiana}
  \city{Lugano}
  \country{Switzerland}
  }
\email{paolo.tonella@usi.ch}

\renewcommand{\shortauthors}{M. Biagiola et al.}

\begin{abstract}
\input{0-abstract}

\end{abstract}

%
%




\maketitle

\input{1-introduction}
\input{2-background}
\input{3-approach}

\input{4-evaluation}

\input{5-discussion}
\input{6-related}
\input{7-conclusion}

\balance
\bibliographystyle{acm}
\bibliography{ms}

\end{document}

%% file: 0-abstract.tex
E2E web test suites are prone to test dependencies due to the heterogeneous multi-tiered nature of modern web apps, which makes  it difficult for developers to create isolated program states for each test case. 
In this paper, we present the first approach for detecting and validating test dependencies present in E2E web test suites. 
Our approach employs string analysis to extract an approximated set of dependencies from the test code. It then filters potential false dependencies through natural language processing of test names. Finally, it validates all dependencies, and uses a novel recovery algorithm to ensure no true dependencies are missed in the final test dependency graph.
Our approach is implemented in a tool called \tool and evaluated on 
the test suites of six open-source web apps. Our results show that \tool can correctly detect and validate test dependencies up to 72\% faster 
than the baseline with the original test ordering in which the graph contains all possible dependencies.
The test dependency graphs produced by \tool enable test execution parallelization, with a speed-up factor of up to 7$\times$. 

%% file: 1-introduction.tex
\section{Introduction}\label{sec:introduction}


Ideally, all tests in a test suite should be independent. 
However, in practice, developers create tests that are \textit{dependent} on each other~\cite{Gyori:2015:RTD:2771783.2771793,Zhang:2014:ERT:2610384.2610404,pradet,Bell:2015:EDD:2786805.2786823,Kappler:2016:FBT:2950290.2983974}. 
%
Test dependency can be informally defined as follows.
Let $T=\langle t_1, t_2, \ldots, t_n\rangle$ be a test suite, where each $t_i$ is a test case, whose index $i$ defines an order relation between test cases that corresponds to the original execution order given by testers. When tests within $T$ are executed in the original order, 
all tests execute correctly.
If the original execution ordering is altered, e.g., by executing $t_2$ before $t_1$, and the execution of $t_2$ fails, we can say that $t_2$ \textit{depends on} $t_1$ for its execution, and that a \textit{manifest test dependency} exists~\cite{Zhang:2014:ERT:2610384.2610404,pradet}.

Test dependencies inhibit the use of test optimization techniques such as test parallelization~\cite{Bell:virtual}, test prioritization~\cite{Rothermel:2001:PTC:505464.505468}, test selection~\cite{Gligoric:2015:PRT:2771783.2771784} and test minimization~\cite{Vahabzadeh:ICSE18}, which all require having independent test cases.
Furthermore, test dependencies can mask program faults and lead to undesirable misleading side-effects, such as tests that pass when they should fail, tests that fail when they should pass~\cite{Zhang:2014:ERT:2610384.2610404,LamZE2015}, or significant test execution overheads~\cite{Bell:2015:EDD:2786805.2786823,Kappler:2016:FBT:2950290.2983974}. 

Web developers frequently use end-to-end (E2E) test automation tools, which 
verify the correct functioning of the application in given test scenarios by means of automated test scripts. Such scripts automate the manual operations that the end user would perform on the web application's graphical user interface (GUI), such as delivering events with clicks, or filling in forms~\cite{STVR:STVR121,Fewster,7381848}.
Unlike traditional unit tests, E2E tests focus on whole business-oriented scenarios, such as logging into the web application, adding items to a shopping cart, and checking out. Such test scenarios go through all tiers and involve all services required to make the whole application work. 
Thus, in web testing, it can be difficult to enforce isolation, as web tests might use the application state which is promptly available from previous test case executions (i.e., a polluted state~\cite{Gyori:2015:RTD:2771783.2771793}).
This creates potential test dependencies due to  read-after-write operations performed on persistent data such as database records or Document Object Model (DOM) fragments that are written by a test $t_i$ and later accessed by a successive test $t_j$ (where $j > i$).


Automated detection of all test dependencies in any given test suite is NP-complete~\cite{Zhang:2014:ERT:2610384.2610404}. As such, researchers have proposed techniques and heuristics that help developers detect an approximation of such dependencies in a timely manner~\cite{Gyori:2015:RTD:2771783.2771793,Zhang:2014:ERT:2610384.2610404,pradet,Bell:2015:EDD:2786805.2786823,Kappler:2016:FBT:2950290.2983974}. 
%
In this work, we focus on automatically detecting test dependencies in E2E web tests.
Existing tools such as DTDetector~\cite{Zhang:2014:ERT:2610384.2610404}, ElectricTest~\cite{Bell:2015:EDD:2786805.2786823} or PRADET~\cite{pradet} are not applicable  because they are based on the extraction of read/write operations  affecting shared data (e.g., static fields) of Java objects. 
Instead, web applications are prone to dependencies due to the persistent data managed on the server-side and the implicit shared data structure on the client-side represented by the DOM. Such dependencies are spread across multiple tiers/services of the web application architecture and are highly dynamic in nature. Hence, existing techniques based on static code analysis are not directly applicable. The web test dependency problem demands for novel approaches that leverage the information available in the web test code and on the client side.


In this paper, we propose a novel test dependency detection technique for E2E web test cases based on \emph{string analysis} and \emph{natural language processing} (NLP). 
Our approach is implemented in a tool called \tool (\textbf{Te}st \textbf{D}epen\-den\-cy \textbf{D}etector), which supports efficient and conservative detection and validation of test dependencies in an E2E test suite using only client-side information, which makes it independent of the server-side technology. 

\tool \textit{extracts} an initial approximated dependency graph (TDG). Then, it \textit{filters} potentially false dependencies in order to speed up the validation process.
Afterwards, it \textit{validates} each dependency by dynamic analysis and it \textit{recovers} any manifest dependency that is potentially missing in the initial and/or filtered graph.

The output of \tool is a validated TDG, which ensures that any test execution schedule that respects its dependencies will not result in any test failure.
In our empirical study on six web test suites, \tool produced the final TDGs 72\% faster than a baseline approach that validates all possible dependencies (57\%, on average). Also, the test suites parallelized by \tool according to the dependencies in the final TDGs achieved a speedup up to 7$\times$ (2$\times$ on average).

Our paper makes the following contributions:
\begin{itemize}[noitemsep]
\item The first test dependency detection approach for web tests. Our approach introduces string analysis (SA) to extract an approximated set of test dependencies, and NLP/SA to filter potential false dependencies. 
\item An algorithm to automatically retrieve all missing dependencies from any given web test dependency graph. 
\item An implementation of our algorithm in a tool named \tool. 
\item An empirical evaluation of \tool on a benchmark of \numberOfSubjects open-source web test suites, comprising \totalNumberOfTests test cases. 
\end{itemize}


%% file: 2-background.tex
\section{Background and Motivation}\label{sec:background}

Ideally, running the tests in a test suite in any order should produce the same outcome~\cite{pradet}. This means tests should deterministically  pass or fail \textit{independently} from the order in which they are executed.
%
A test dynamically alters the state of the program under test in order to assert its expected behaviour.
In practice, some tests  fail to undo their effects on the program's state after their execution, which can pollute any shared state~\cite{Gyori:2015:RTD:2771783.2771793} in tests executed subsequently.
In the web domain, testers perform end-to-end (E2E) testing of their applications~\cite{Berner,Fewster} by creating test cases using test automation tools such as Selenium WebDriver~\cite{selenium}. 
Such tests 
 consist of (1)~actions that simulate an end-user's interactions with the application, and (2)~assertions on information retrieved from the web page to verify the expected behaviour. Unlike unit testing, in which tests target specific class methods, web tests simulate E2E user scenarios, and therefore the program state that persists across test case executions might be left polluted, causing test failures if tests are reordered.

\head{Motivating Example}
\autoref{table:test-suite} lists six E2E Selenium WebDriver tests 
for the Claroline web application~\cite{WCRE}, one of the  subject test suites used in our evaluation. 

\input{table-test-suite}

\autoref{fig:ab1} shows dependencies between tests $t_1$ and $t_2$. 
The test case \texttt{addUserTest} logs in to the application with the administrator credentials (lines~3--5), it navigates to the create user page (lines~6--7), it creates a new user account having username \texttt{user001} by filling in and submitting the appropriate form (lines~8--12), and it finally verifies that a message is correctly displayed (line~13). 

The execution of \texttt{addUserTest} pollutes the state of the web application, which is used by the subsequent test    \texttt{searchUserTest} to search for the same user \texttt{user001} created by \texttt{addUserTest} (line~22). Thus, the shared input data \texttt{user001} might reveal a potential dependency between the tests (see highlighted inputs in \autoref{fig:ab1}).

To make these two tests independent and avoid polluted program states, a tester, for instance, should (1)~delete the user \texttt{user001} created in \texttt{addUserTest}, to clean the polluted program state, and (2)~re-create the same user (or a different one) in \texttt{searchUserTest}. 

In practice, however, testers re-use states created by preceding tests to avoid test redundancy, higher test maintenance cost and increased test execution time~\cite{2016-Leotta-Advances}.
In doing so, they also enforce pre-defined test execution orders, which in turn inhibit utilizing test optimization techniques such as test prioritization~\cite{Rothermel:2001:PTC:505464.505468}.

\input{motivation-example-and-graph}

\input{approach-diagram}

\head{Test Dependency Graph}
The dependencies occurring between tests can be  represented in a test dependency graph ($TDG$)~\cite{pradet}. $TDG$ is a directed acyclic graph in which nodes represent test cases and edges represent dependencies. $TDG$ contains an edge from a test  $t_2$ to a test  $t_1$ if $t_2$ depends on $t_1$ for its execution (notationally, $t_2 \rightarrow t_1$). 

\autoref{fig:ab2} illustrates the actual test dependency graph ($TDG$) for the test suite of \autoref{table:test-suite}. 
For example, $TDG$ contains an edge from \texttt{searchCourseTest} to \texttt{addCourseTest} because \texttt{searchCourseTest} requires the execution of \texttt{addCourseTest} to produce the expected result (in other words, \texttt{addCourseTest} \textit{must be executed before} \texttt{searchCourseTest} in order for it to succeed).
Multiple test dependencies can also occur. For instance, \texttt{enrolUserTest} depends on both \texttt{addUserTest} and \texttt{addCourseTest} for its correct execution.

In order to be useful, $TDG$ should contain all \textit{manifest dependencies}, i.e., dependencies that do cause tests to fail if violated, while retaining the minimum number (or none) of \textit{false dependencies}.

One possible application of a $TDG$ that contains only manifest dependencies is test suite parallelization. For instance, if we traverse the graph of \autoref{fig:ab2} and extract the subgraphs reachable from each node with zero in-degree (in our example, there is only \texttt{addUserTest}), we can identify subsets of tests that can be be executed in parallel with the others. 
In our example, four parallel test suites are possible: 
\{ $\langle t_1, t_2 \rangle$, 
$\langle t_1, t_3 \rangle$, 
$\langle t_1, t_4, t_6 \rangle$, 
$\langle t_4, t_5 \rangle$ \}.

%% file: table-test-suite.tex
\begin{table}[h]
\setlength{\tabcolsep}{5pt}
\renewcommand{\arraystretch}{1}
\footnotesize
\centering
\caption{Test cases for Claroline, numbered according to their test execution order~\cite{WCRE}.}
\label{table:test-suite}
\begin{tabular}{lll}
\toprule
\textbf{Test} & \textbf{Name} & \textbf{Description} \\
\midrule
$t_1$ & \texttt{addUserTest} & The admin creates a new user account. \\
$t_2$ & \texttt{searchUserTest} & The admin searches for the newly created user. \\
$t_3$ & \texttt{loginUserTest} & The newly created user logs in to the application. \\
$t_4$ & \texttt{addCourseTest} & The admin creates a new course. \\
$t_5$ & \texttt{searchCourseTest} & The admin searches for the newly created course. \\
$t_6$ & \texttt{enrolUserTest} & The user enrols herself in the course. \\
\bottomrule
\end{tabular}
\end{table}

%% file: motivation-example-and-graph.tex
\def\tikzmark#1{\tikz[remember picture,overlay]\node[yshift=2pt](#1){};}

\begin{figure}[t]
\centering

\begin{lstlisting}[]
@Test
public void addUserTest() {
	driver.findElement(By.id("login")).sendKeys(<@\hlsmooth{"admin"}@>);
	driver.findElement(By.id("password")).sendKeys(<@\hlsmooth{"admin"}@>);
	driver.findElement(By.xpath("//button")).click();
	driver.findElement(By.linkText("Platform administration")).click();
	driver.findElement(By.linkText("Create user")).click();
	driver.findElement(By.id("lastname")).sendKeys(<@\hlsmooth{"Name001"}\tikzmark{A}@>);
	driver.findElement(By.id("firstname")).sendKeys(<@\hlsmooth{"Firstname001"}\tikzmark{B}@>); 
	driver.findElement(By.id("username")).sendKeys(<@\hlsmooth{"user001"}\tikzmark{C}@>);
	driver.findElement(By.id("password")).sendKeys(<@\hlsmooth{"password001"}@>);
	driver.findElement(By.id("password_conf")).sendKeys(<@\hlsmooth{"password001"}@>);
	assertEquals("The new user has been created", driver.findElement(By.xpath("//*[@id='claroBody']")).getText());
	driver.findElement(By.id("logout")).click();
}
@Test
public void searchUserTest() {
	driver.findElement(By.id("login")).sendKeys(<@\hlsmooth{"admin"}@>);
	driver.findElement(By.id("password")).sendKeys(<@\hlsmooth{"admin"}@>);
	driver.findElement(By.xpath("//button")).click();
	driver.findElement(By.linkText("Platform administration")).click();
	driver.findElement(By.id("search_user")).sendKeys(<@\hlsmooth{"user001"}\tikzmark{F}@>);
	driver.findElement(By.cssSelector("input[type='submit']")).click();
	assertEquals(<@\hlsmooth{"Name001"} \tikzmark{D}@>,driver.findElement(By.id("L0")).getText());
	assertEquals(<@\hlsmooth{"Firstname001"}\tikzmark{E}@>,driver.findElement(By.xpath("//td[3]")).getText());
	driver.findElement(By.id("logout")).click();
}
\end{lstlisting}

\tikzset{Empharrow/.style={gray,thick,dotted,->,arrows={Bar-Triangle[width=3pt]}}}

\begin{tikzpicture}[remember picture,overlay]
\draw[Empharrow](A)--+(0cm,0)--(D);
\draw[Empharrow](B)--+(0cm,0)--(E);
\draw[Empharrow](C)--+(0cm,0)--(F);
\end{tikzpicture}


\caption{Two dependent E2E web tests for the Claroline web application. Potential dependencies due to shared input data are highlighted.}
\label{fig:ab1} 
\end{figure}

\begin{figure}
\centering

\includegraphics[trim=0cm 20cm 0cm 0cm, clip=true, scale=0.125]
{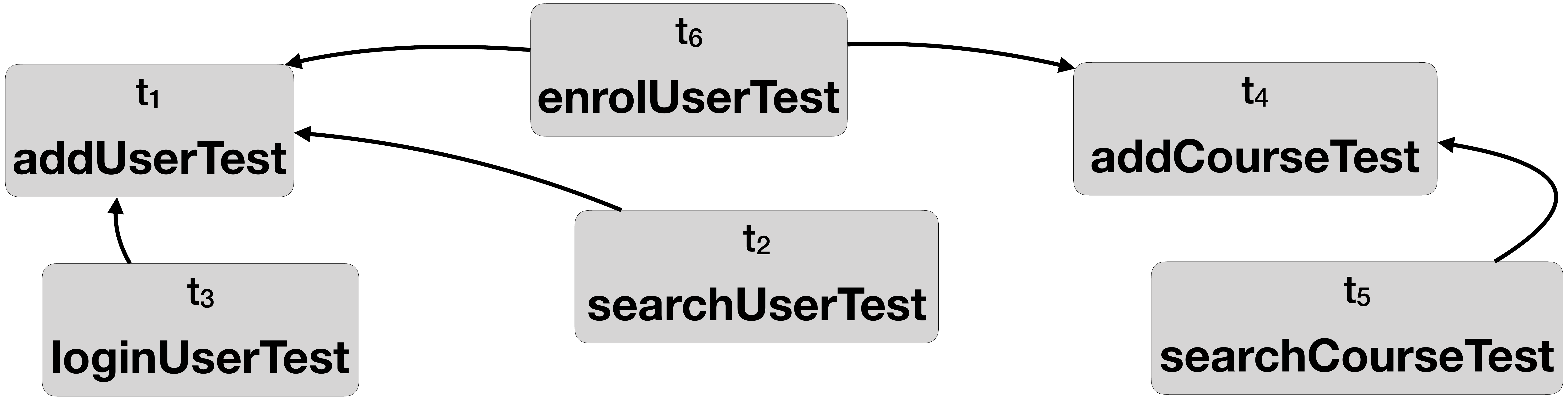}

\caption{The test dependency graph for the test suite of \autoref{table:test-suite}. Solid edges represent \textit{manifest} dependencies, namely, dependencies that result in a different test result if they are not respected.}
\label{fig:ab2}
\end{figure}


%% file: approach-diagram.tex
\begin{figure*}[t]
\centering

\includegraphics[trim=0cm 20cm 0cm 0cm, clip=true,scale=0.20]
{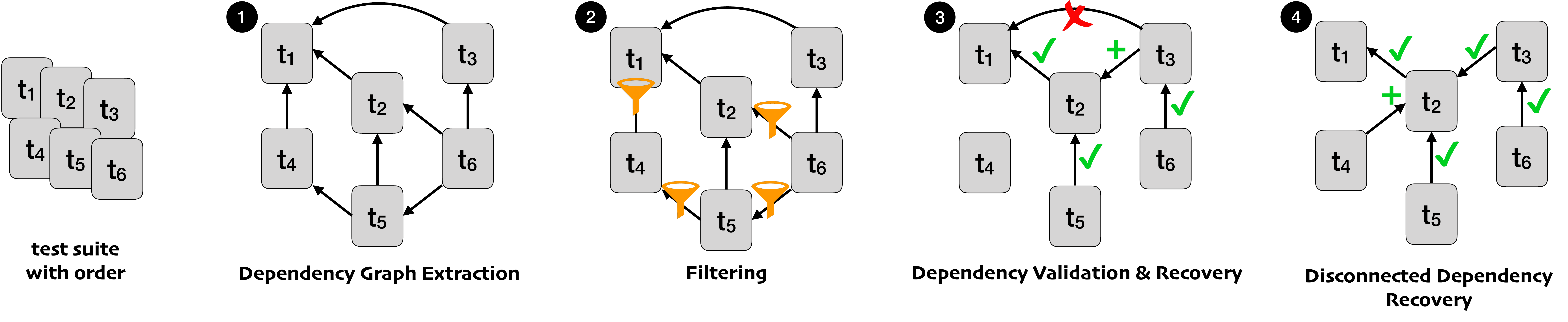}

\caption{Our overall approach for web test dependency detection, validation and recovery.} 
\label{fig:approach} 
\end{figure*}

%% file: 3-approach.tex
\section{Approach
}\label{sec:approach}

The goal of our approach is to automatically detect the occurrence of dependencies among web tests. 
Detecting  dependencies in E2E web tests is particularly challenging due to the stack of programming languages and technologies involved in the construction of modern web applications, e.g., HTML, CSS, JavaScript on the client side; PHP, Java or JavaScript on the server-side, and back-end layers through Restful APIs or databases. 

Our insight is that by analyzing the input data used in test cases (highlighted in \autoref{fig:ab1}), we can obtain clues about potential test dependencies caused by shared polluted states. 
For example, the input string \texttt{user001} used at line~10 in \texttt{addUserTest} is a \textit{write operation} of persistent data. The same input string is used at line~22 in \texttt{searchUserTest} as a \textit{read operation} of persistent data.
We conjecture that such \emph{read-after-write} connections on persistent data could indicate potential test dependencies. Given this insight, our approach focuses on read-after-write relationships on persistent data, defined as follows.

\begin{defn}[\textbf{Persistent Read-After-Write (\praw) Dependency}]
Two test cases $t_1$ and $t_2$ executed in this order in the original test suite are subject to a \praw dependency if $t_1$ performs an operation that writes some information $S_i$ into the persistent state of the web application and $t_2$ performs an operation that reads $S_i$ from the persistent state of the web application.
\end{defn}

Examples of the write operations in $t_1$ include creating, updating or deleting a record in a database, or creating a DOM element on the webpage. Examples of read operations in $t_2$ are reading the same record from the database, or accessing the newly created DOM element on the webpage.

\autoref{fig:approach} illustrates our overall approach, which requires a web test suite as input, along with a predefined test execution order.  Overall, our approach \ding{182}~computes an initial approximated test dependency graph, 
\ding{183}~filters out potential false dependencies,  \ding{184}~dynamically validates all dependencies in the graph while recovering any missing dependencies, and, finally, \ding{185}~handles missing dependencies affecting independent nodes possibly resulting from the previous validation step. 



Next, we describe each step of our approach.

\subsection{Dependency Graph Extraction}\label{sec:extraction}

In the first step, from the input test suite, our approach computes an initial test dependency graph representing an approximated set of candidate dependencies. 
This can be conducted in different ways, as  described below.

\subsubsection{Original Order Graph Extraction}\label{sec:approach-conservative} 

A baseline approach consists of connecting all pairwise combinations of tests according to the original order. This results in a directed graph in which every pair of distinct nodes is connected by a unique pair of edges, so as to establish a dependency relation between each test and all the others that are executed before it. 
If $n$ is the number of test cases in the test suite, the graph contains $\frac{n(n-1)}{2}$ edges (e.g., dependencies). 

Since the time to validate a dependency graph increases with the number of dependencies in the graph, heuristics can be used to reduce the size of the graph by removing edges that are less likely to be manifest dependencies. 
To that end, we propose an approach that leverages a fast static \textit{string analysis} of the input data present in the tests,  to construct a smaller initial test dependency graph. 

\subsubsection{Sub-Use String Analysis Graph Extraction}\label{sec:approach-sa}

\autoref{algorithm:string-analysis} describes our dependency graph extraction based on sub-use-chain relations. A sub-use relation  consists of a submission (\textit{sub}) of input data $i$ and all the following \textit{uses} that submitted value $i$.

\input{algo-string-analysis}

Starting from the first test case according to the original test suite  order, \autoref{algorithm:string-analysis} first retrieves the set $\mathcal{S}$ of input values \textit{submitted} by the test, like values inserted into input fields by input-submitting actions such as the \texttt{sendKeys} methods (line~4). 

Second, the algorithm considers each test case $t_f$ following $t$ and searches for any input value in the set $\mathcal{S}$, which is \textit{used} in any statement of the current test case $t_f$ (line~7), and adds them to the set of \textit{used} values $\mathcal{U}$. 
If at least one string value is found (i.e., a \textit{sub-use} chain), a candidate \praw dependency between $t$ and $t_f$ is created by adding the edge $t_f \rightarrow t$, labelled with each retrieved string value (line~9), to the test dependency graph (line~10).

The internal loop of \autoref{algorithm:string-analysis} (lines 6--12) searches for the input values indistinctly in any test action because in web tests there is no clear distinction between read and write statements. For instance, in \autoref{fig:ab1}, the \texttt{sendKeys} action at line~10 is used to \textit{write} persistent information into the application (e.g., in a database). However, the same \texttt{sendKeys} action, at line~22, identifies an action with a \textit{read} connotation, because the string value \texttt{user001}  is used to search for a specific persistent information in  the web application. 

Let us consider the source code of the two test cases in \autoref{fig:ab1}, namely \texttt{addUserTest} and \texttt{searchUserTest}. 
From the first test \texttt{addUserTest}, the algorithm extracts the set $\mathcal{S} = \{$\texttt{admin}, \texttt{Name001}, \texttt{Firstname001}, \texttt{user001}, \texttt{password001}$\}$ because \texttt{sendKeys} is the only input-submitting action. 
Then, the string values in $\mathcal{S}$ are looked up in the subsequent test \texttt{searchUserTest}, producing the set of used values $\mathcal{U}=\{$\texttt{admin}, \texttt{Name001}, \texttt{Firstname001}, \texttt{user001}$\}$.
Being $\mathcal{U}$ not empty, the algorithm creates a candidate test dependency between \texttt{searchUserTest} and \texttt{addUserTest}. 

\subsection{Filtering}\label{sec:approach-filtering}

The second step of our approach applies a filtering process to remove potential false \praw dependencies. 
The filtering is performed to speed up the subsequent validation step, which requires in-browser test execution, and therefore can be computationally expensive for graphs with numerous candidate test dependencies. 
%
Finding an effective filtering technique is, however, challenging. A \textit{loose} filter might remove a few false dependencies, whereas a \textit{strict} filter might mistakenly remove manifest dependencies, which would need to be recovered at a later stage. 

In this work, we propose two novel test dependency filtering techniques based on (1)~dependency-free values, and (2)~Natural Language Processing (NLP). 


\subsubsection{Dependency-free String Value Filtering}\label{sec:approach-sa-filtering}

We analyze the frequency of string input values used in the test suite to filter potential false \praw dependencies.

Let us consider the test dependency graph depicted in \autoref{fig:graph-SA}, obtained by applying our sub-use string analysis graph extraction (\autoref{sec:approach-sa}) to the motivating example test suite (\autoref{sec:background}).

\input{graph-SA}

The set of dependencies \{ $t_5 \rightarrow t_2$, $t_5 \rightarrow t_1$, $t_4 \rightarrow t_1$ \} represents instances of \textit{false} candidate \praw dependencies. The edges between these test cases are only due to the same login input data used by the tests---i.e., \texttt{admin}---a default user created during the installation, for which no test case must be executed to create it.

Existing techniques~\cite{Zhang:2014:ERT:2610384.2610404,pradet} refer to such cases as \textit{dependency-free values}, i.e., if the test dependency graph includes dependencies that are shared across multiple (or all) test cases, these likely-false dependencies could be filtered out. 

%

However, in principle, these assumptions might not hold in all cases, as occurrence frequency alone is not conclusive for safe filtering. 
Our dependency-free string value filtering computes a ranked list of frequently occurring strings and asks the developer to either confirm or discard them (if a string value occurs in all test cases, the corresponding dependency is automatically filtered).

In our example of \autoref{fig:graph-SA}, our approach computes the frequencies of all strings, presents it to the the tester, who, for instance, may decide to filter the dependencies due to the \texttt{admin} string, hence removing $t_5 \rightarrow t_2$, $t_5 \rightarrow t_1$, $t_4 \rightarrow t_2$.

\subsubsection{NLP-based Filtering}\label{sec:approach-nlp-filtering}

Developers often use descriptive patterns for test case names, which summarize the operations performed by each test. 
Giving a descriptive name to a test case has several advantages such as enhanced \textit{readability} (i.e., it becomes easier to understand what behaviour is being tested) and \textit{debugging} (i.e., when a test case fails, it is easier to identify the broken functionality). For instance, Google recommends test naming conventions~\cite{google-junit-test-names,google-scenario-test-names} in which \textit{unit tests} need to be named with the method being tested (a verb or a verb phrase, e.g., \texttt{pop}) and the application state in which the specific method is tested (e.g., \texttt{EmptyStack}). 
%
%
For \textit{behaviour-based} tests such as E2E tests~\cite{google-scenario-test-names}, the guidelines propose a naming convention that includes the test scenario (e.g., \texttt{invalidLogin}) and the expected outcome (e.g., \texttt{lockOutUser}). 

Therefore, our second filtering mechanism consists of using Natural Language Processing (NLP) to analyze test case names and classify them into two classes, namely, \emph{read} or \emph{write}. 
Then, based on such classification, non-\praw dependencies, such as a ``read'' test being dependent on another ``read'' test, are discarded from the test dependency graph.

Our approach uses a \textit{Part-Of-Speech tagger} (POS) to classify each token (i.e., word) in a tokenized test name as noun, verb, adjective, or adverb. 
In particular, our approach uses the verb from the test case name as the part of speech that conveys the nature of the test operation, and uses it to classify each test into \textit{read} or \textit{write} classes. 
Our approach relies on two groups of standardized R/W verbs, namely CRUD operations---\textit{Create}, \textit{Read}, \textit{Update} and \textit{Delete})~\cite{crud}---in which the \textit{Read} operation is pre-classified as read whereas the other three are pre-classified as write. 

Our approach uses POS to extract the first verb from each test name and then computes the semantic similarity~\cite{wordnet-similarity} (specifically, the \textit{WUP} metrics~\cite{wup}) between the extracted verb and each verb in the pre-classified read/write groups.
The similarity score quantifies how much two concepts are alike, based on information contained in the \textit{is--a} hierarchy of \textit{WordNet}~\cite{wordnet}. 
After computing all similarity scores, our approach classifies the extracted verb to the group having the maximum similarity score. 
In case of ties (e.g., the verb has the same similarity score for both the read and write classes), or in case no verbs are found, our approach does not perform any assignment and the dependency is not filtered. Our classification of read/write verbs achieved a precision of 80\% and a recall of  94\% on our experimental subjects.

In this work, we propose and evaluate three NLP configurations. 

\head{Verb only (NLP Verb)}
Our first NLP filtering configuration considers only the \textit{verb of the test case name}. 
Given a dependency $t_y \rightarrow t_x$, our approach extracts the verb from both  $t_y$ and $t_x$, and classifies them either as read or write. 
Then, it filters (1)~the read-after-read (RaR) dependencies, in which both $t_y$ and $t_x$ have verbs classified as read, and (2)~the write-after-read (WaR) dependencies, where $t_y$ has a write-classified verb whereas $t_x$ has a read-classified verb. 
All other types of dependencies, such as read-after-write (RaW) and WaW (write operations in web applications often also involve reading existing data), are retained. 

The dependency \texttt{searchCourseTest} $\rightarrow$ \texttt{searchUserTest} (\autoref{fig:graph-SA}) is filtered because it is classified as RaR, with \texttt{search} being the read-classified verb.  Conversely, the edge \texttt{searchUserTest} $\rightarrow$ \texttt{addUserTest} is retained since it is classified as RaW, being \texttt{search} and \texttt{add} the read/write verbs, respectively. The word \texttt{Test} is considered a \textit{stop word} and removed before the NLP analysis starts.

\head{Verb and direct object (NLP Dobj)}
The second configuration considers the \textit{direct object} the verb refers to. 
Given a set of test cases, our approach uses a dependency parser to analyze the grammatical structure of a sentence, extract the direct object from each test name, and construct a set of ``dobject'' dependencies. 

RaR and WaR dependencies are filtered as described in the previous  NLP Verb case. Differently, RaW and WaW dependencies are filtered only if the direct objects of two verbs appearing in two tests $t_y$ and $t_x$ are different. 
The intuition is that the two tests may perform actions on different persistent entities of the web application, if the involved direct objects in the test names are different.
For example, in \autoref{fig:graph-SA}, the RaW dependency \texttt{searchCourseTest} $\rightarrow$ \texttt{addUserTest} is filtered because the two involved direct objects, \texttt{Course} and \texttt{User}, are different.

\head{Verb and nouns (NLP NOUN)}
Our third configuration takes into account all entities of type \textit{noun} contained in the test names. 
When the test name includes multiple, different entities, analyzing only the direct object may not be enough to make a safe choice. 
For instance, in our subject Claroline, the analysis of the direct object would erroneously filter the manifest dependency \texttt{addCourseEventTest} $\rightarrow$ \texttt{addCourseTest}, because it is a WaW and the two direct objects \texttt{Event} and \texttt{Course} are different. 
However, there is an implicit relation between the direct object \texttt{Event} and the \texttt{Course} object it refers to.
Thus, the dependency with \texttt{addCourseTest} should be retained.

Again, RaR and WaR dependencies are filtered as described in the NLP Verb case. Here, RaW and WaW dependencies are filtered only if the two tests involved in a dependency have no noun in common. As such, the manifest dependency \texttt{addCourseEventTest} $\rightarrow$ \texttt{addCourseTest} would not be filtered in this configuration, because of the shared name \texttt{Course}.
 



\subsection{Dependency Validation and Recovery}\label{sec:approach-da}


Given a test dependency graph $TDG$, the overall dynamic dependency validation procedure works according to the  iterative process proposed by Gambi et al.~\cite{pradet}. 
The approach executes the tests according to the original order to store the expected outcome. Next, it selects a target dependency according to a source-first strategy in which tests that are executed later in the original test suite are selected \textit{first} (i.e., $t_3 \rightarrow t_2$ would be selected before $t_2 \rightarrow t_1$). 

To validate the target dependency, tests are executed out of order, i.e., 
a test schedule in which the target dependency is \textit{inverted} is computed and executed. 
If the result of the test execution differs from the expected outcome, the target dependency is marked as \textit{manifest}, because the failure was due to the inversion. Otherwise, the target dependency is removed from $TDG$. The process iterates until all dependencies are either removed or marked as manifest.

The dynamic validation procedure described above works correctly under the assumption that the initial TDG contains all manifest dependencies (as the original order graph \autoref{sec:approach-conservative}). 
In our approach, the filtering techniques applied in the previous step may be not conservative. 
Therefore, differently from existing techniques~\cite{pradet}, our approach features a dynamic dependency recovery mechanism that retrieves all \emph{missing} dependencies. To the best of our knowledge, this is the first dependency validation algorithm that also includes dynamic dependency recovery. 

\input{algo-within-validation-recovery}

\head{Recovering Missing Dependencies}
\autoref{algorithm:within-validation-recovery} takes a partially-validated $TDG$. For each failing test schedule in which a target dependency is inverted, it checks whether the failure is due to a missing dependency in the dependency graph.

More specifically, \autoref{algorithm:within-validation-recovery} takes the target dependency and computes a schedule in which the target dependency \textit{is not} inverted (line~1). If the execution of such schedule complies with the expected outcome, our approach considers the test failure due to the dependency inversion and marks the dependency as a manifest. On the contrary, if one or more tests fail also in the schedule without inversion (line~4), our approach assumes that one or more dependencies are missing and need to be recovered. 

To do so, the algorithm takes the first failing test and retrieves the preceding test cases that were not executed in the schedule (line~6). Those tests are all candidate manifest dependencies for the failed test. The algorithm connects the failed test case with each such preceding test and adds those dependencies to the graph (lines~8--9). The graph obtained this way contains all newly added candidate manifest dependencies that still need to be validated. 




\input{example-recovery.tex}

Let us take as example \autoref{fig:example-recovery}.A, in which  $t_4$ has a \textit{missing manifest dependency} on $t_2$, and $t_3$ does not modify the application state. According to the source-first strategy, the validation selects the dependency $t_4 \rightarrow t_3$ . The schedule computed for such dependencies is $\langle t_4\rangle$, in which $t_4$ fails because $t_2$ is not executed. Then, our algorithm starts retrieving the missing dependency by computing a schedule in which $t_4 \rightarrow t_3$ is not inverted, $\langle t_3, t_4\rangle$, in which $t_4$ fails again for the same reason. The recovery procedure concludes that there is at least one missing dependency, and connects $t_4$ with both $t_1$ and $t_2$ (\autoref{fig:example-recovery}.B), i.e., the only candidate manifest dependencies.

In \autoref{fig:example-recovery}.C the dependency $t_4 \rightarrow t_3$ is selected again. This time, the computed schedule is $\langle t_1, t_2, t_4\rangle$, in which none of the tests fail. Therefore, the dependency is marked as false and removed. The next selected dependency is $t_4 \rightarrow t_2$, for which the schedule $\langle t_1, t_4\rangle$ is computed. The test $t_4$ fails because $t_2$ is not executed. To check if the failure is due to a missing dependency, our algorithm computes the test schedule $\langle t_1, t_2,t_4\rangle$, in which none of the tests fail. Our algorithm concludes that $t_4 \rightarrow t_2$ is a manifest dependency and \textit{recovers} it. The validation iterates over the other dependencies in the same way and  outputs the final $TDG$ (\autoref{fig:example-recovery}.D), where the initially missing dependency has been recovered.

\subsection{Disconnected Dependency Recovery}\label{sec:ind-node-recovery}
The previous validation step~\ding{184} can produce disconnected components in the $TDG$. Missing dependencies involving tests in disconnected components require a separate treatment.
Two cases can occur (1)~tests with no outgoing edges (zero out-degree),\footnote{First test $t_1$ excluded} and (2)~\textit{isolated} tests, i.e., tests having neither incoming nor outgoing edges (zero in- and out-degree). 

The former case occurs when a false dependency, removed during the validation, \textit{shadows} a missing dependency. In such cases, disconnected components of $TDG$, including potentially missing dependencies, are created as a result of the validation.

\autoref{fig:example-independent}.A illustrates an example: the manifest dependency $t_4 \rightarrow t_1$ is missing in the initial dependency graph. Let us suppose that $t_3$ does not change the state of the application when executed, and therefore, its execution does not influence the execution of any other successive test in the original order. 
The algorithm selects first the dependency $t_4 \rightarrow t_3$, it produces the schedule $\langle t_4\rangle$, in which the test fails since $t_1$ is not executed. 
Our approach checks if the failure is due to a missing dependency. When the dependency is \textit{not inverted}, the computed schedule is $\langle t_1$, $t_3$, $t_4\rangle$, in which none of the tests fail. Hence, our algorithm concludes that $t_4 \rightarrow t_3$ is a manifest dependency and no dependency recovery takes place. 
In the next step, the algorithm validates the dependency $t_3 \rightarrow t_1$, which is removed because $t_3$ can execute successfully without $t_1$. 
The dependency graph produced by dependency validation algorithm is illustrated in \autoref{fig:example-independent}.B. In the isolated subgraph $t_4 \rightarrow t_3$ the schedule $\langle t_3$,$t_4\rangle$ results in a failure of $t_4$. 
Indeed, the dependency $t_4 \rightarrow t_1$ was not captured by the recovery algorithm because \textit{the false dependency $t_3 \rightarrow t_1$  \textit{shadowed} the absence of the manifest dependency $t_4 \rightarrow t_1$}.

\autoref{fig:example-independent}.A also illustrates how our approach handles isolated tests. In this example, $t_2$ is an isolated node. Let us suppose that $t_2$ has a manifest dependency on $t_1$ ($t_2 \rightarrow t_1$), which is missing in the initial $TDG$ because it is either not captured, or because it is wrongly filtered out in the second step of our approach. 
Therefore, the validation step would produce the $TDG$ shown in \autoref{fig:example-independent}.B, in which there is no chance to check whether $t_2$ executes successfully in isolation. In fact, $t_2$ is not part of any test schedule that can be generated from $TDG$, regardless of any possible dependency inversion. For this reason, a further recovery step is required once the validation is completed.

\input{example-independent.tex}

\input{algo-post-validation-recovery}

\autoref{algorithm:post-validation-recovery} handles the recovery of missing dependencies within disconnected components. The algorithm retrieves all isolated nodes and zero out-degree nodes (line~3) and executes each of them in isolation (line~5). 
For each failing test, the algorithm connects it with all preceding tests according to the initial test suite order (line~8). 
Otherwise, if a test is not isolated and executes successfully (line~9), the algorithm takes all schedules that contain that test and execute them (lines $12$--$18$). If a test in those schedules fails, the algorithm connects it with all the preceding ones (line~16). 

Finally, for each dependency found and added to $TDG$ during the disconnected dependency recovery step, the dependency validation procedure must be re-executed.

Given the graph in \autoref{fig:example-independent}.B, \autoref{algorithm:post-validation-recovery} executes $t_2$ in isolation, which fails, thus the dependency $t_2 \rightarrow t_1$ is added. Moreover, in  \autoref{fig:example-independent}.B, there is only one schedule that involves $t_3$, namely $\langle t_3$,$t_4\rangle$. In this schedule $t_4$ fails, hence our approach adds the dependencies $t_4 \rightarrow t_2$ and $t_4 \rightarrow t_1$ (\autoref{fig:example-independent}.C). Next, the added dependencies are validated (\autoref{fig:example-independent}.D), and the final graph is produced, where all initially missing manifest dependencies have been successfully recovered (\autoref{fig:example-independent}.E).


To conclude, our validation and recovery algorithms makes sure that (1)~newly added dependencies are themselves validated, (2)~false dependencies are removed in the final $TDG$. 
Indeed, a node in the final $TDG$ can be either (i)~connected (i.e., in-degree~$> 0$ and out-degree~$> 0$), (ii)~without outgoing edges (i.e., in-degree~$> 0$ and out-degree~$= 0$) or (iii)~isolated (i.e., in-degree~$=$~out-degree~$= 0$). 


\subsection{Implementation}\label{sec:implementation}

We implemented our approach in a tool called \tool (\textbf{Te}st \textbf{D}epen\-den\-cy \textbf{D}etector). 
The tool is written in Java, and supports Selenium WebDriver web test suites written in Java. However, our overall approach is  general because it can be applied to 
test suites developed using other programming languages or web testing frameworks. 
\tool expects as input the path to a test suite and performs the string analysis by parsing the source code of the tests by using  \textit{Spoon} (version $6.0.0$)~\cite{spoon}. Our NLP module adopts algorithms available in the open-source library \textit{CoreNLP} (version $3.9.2$)~\cite{core-nlp}. 
The output of  \tool is a list of manifest dependencies extracted from the final validated TDG.

%% file: algo-string-analysis.tex
\begin{algorithm}[h!]
	
\DontPrintSemicolon
\scriptsize

\SetKwInOut{Input}{Input}
\SetKwInOut{Output}{Output}
\Input{$T_o$: test suite in its original order $o$, $\mathcal{I}$: set of input-submitting actions}
\Output{$TDG$: test dependency graph with candidate dependencies to be validated}

$TDG$ $\gets$ $\emptyset$\;
$T_a$ $\gets$ $T_o$ \Comment*{tests to analyze}

\ForEach{$t$ in $T_o$}{
	$\mathcal{S} \gets$ \textsc{getInputValues}($t$, $\mathcal{I}$) \Comment*{$\mathcal{S}$ is the set of input values submitted by $t$}
	$T_a$ $\gets$ $T_a - \{t\}$\;
	\ForEach{$t_f$ in $T_a$}{
		$\mathcal{U}$ $\gets$ \textsc{findUsedValues}($t_f$) $\cap\ \mathcal{W}$ \Comment*{$\mathcal{U}$ is the set of input values used by $t_f$} 
		\If{ $\mathcal{U}$ $\neq \emptyset$}{
		\textit{depToAdd} $\gets$ $(t_f \xrightarrow[]{\mathcal{U}} t)$ \Comment*{candidate manifest dependency}
			$TDG \gets TDG \cup \{$ \textit{depToAdd} $\}$ 
		}
	}
}
 \caption{Sub-use string analysis graph extraction}
 \label{algorithm:string-analysis}
\end{algorithm}

%% file: graph-SA.tex
\begin{figure}[t]
\centering

\includegraphics[trim={0cm 0cm 0cm 0cm},clip,width=0.9\linewidth]
{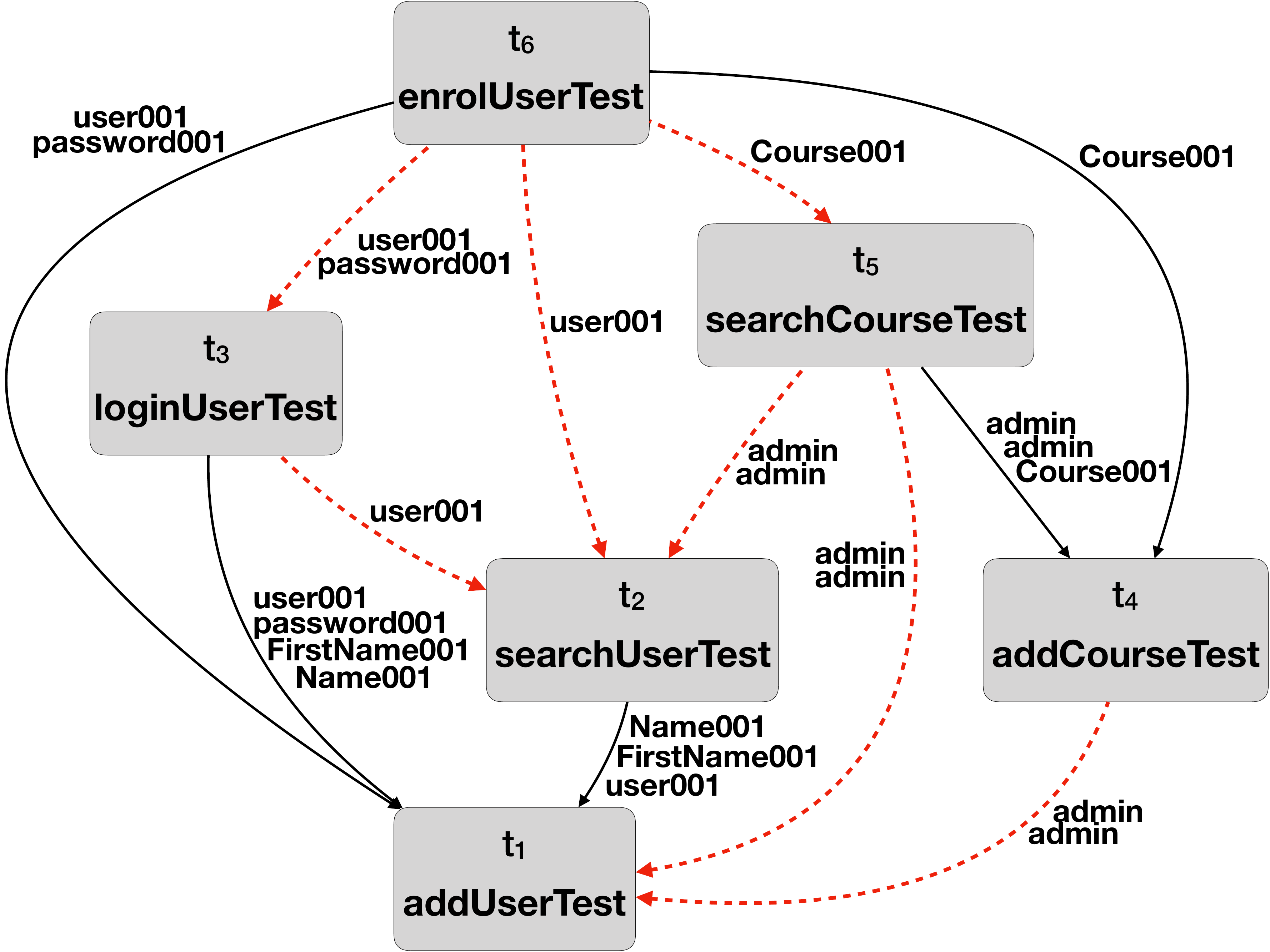}

\caption{Dependency-free string-based \praw filtering. Solid black edges represent manifest dependencies whereas dashed red edges represent false dependencies.} 
\label{fig:graph-SA} 
\end{figure}

%% file: algo-within-validation-recovery.tex
\begin{algorithm}[t]
	
	\DontPrintSemicolon
	\scriptsize
	
	\SetKwInOut{Input}{Input}
	\SetKwInOut{Output}{Output}
	\SetKwInOut{Require}{Require}
	\Input{$T_o$: test suite in its original order $o$ \\
	$TDG$: test dependency graph \\ 
	\textit{targetDep}: dependency selected for validation \\ 
	\textit{expResults}: results of executing $T_o$ \\ 
	\textit{execResults}: results of a test schedule in which \textit{targetDep} is inverted}
	\Output{$TDG$: updated test dependency graph with missing dependencies recovered}
	\Require{\textit{expResults} $\neq$ \textit{execResults}, i.e., \textit{targetDep} is manifest} 
		
		\textit{schedule} $\gets$ \textsc{computeTestScheduleWithNoInversion}($TDG$, \textit{targetDep}) \; 
		\textit{execResults} $\gets$ \textsc{executeTestSchedule}(\textit{schedule})\;
		\textit{failedTest} $\gets$ \textsc{getFirstFailedTest}(\textit{expResults}, \textit{execResults})\;
		\If{failedTest $\neq$ null}{ 
		
		\HiLi{/* Failure due to a missing dependency; get all tests before failedTest. */}
		
		\textit{depCandidates} $\gets$ \textsc{getDepCandidates}$(T_o$, \textit{schedule}$)$ \; 
			\ForEach{depCandidate $\in$ depCandidates}{
				\textit{depToAdd} $\gets$ $\langle$ \textit{failedTest} $\rightarrow$ \textit{depCandidate} $\rangle$ \;
				$TDG \gets TDG \cup \{\textit{depToAdd}\}$
			}
		} 
	\caption{Recovery algorithm}
	\label{algorithm:within-validation-recovery}
\end{algorithm}

%% file: example-recovery.tex
\begin{figure}[h!]
\centering

\includegraphics[trim={0cm 20cm 0cm 0cm},clip,width=\linewidth]
{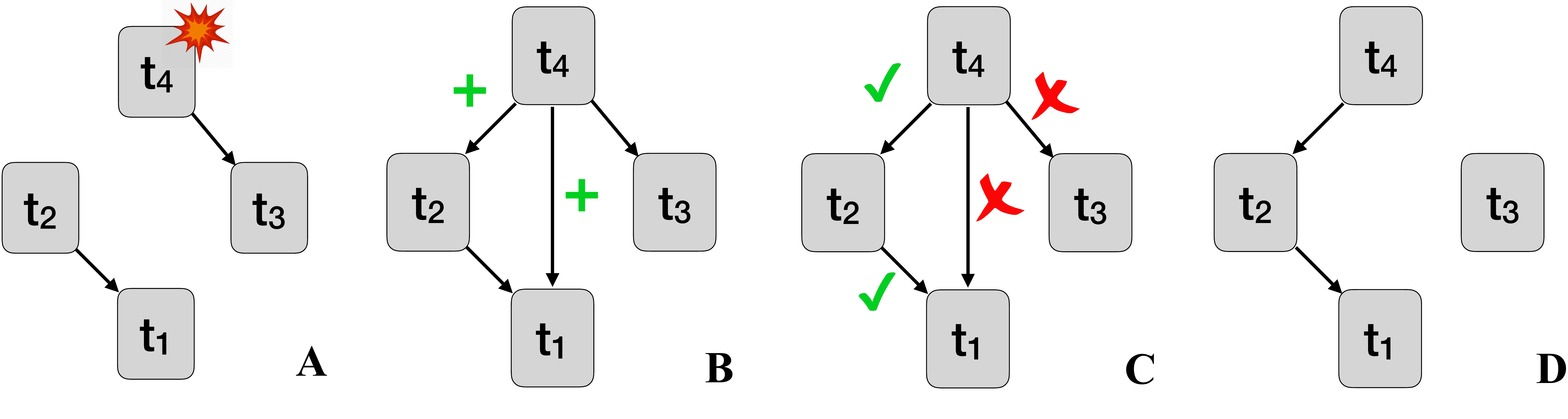}

\caption{Recovery of missing dependencies. (A) $t_4 \rightarrow t_3$ is selected, $t_4$ fails because $t_4 \rightarrow t_2$ is missing. (B) recovery procedure adds candidate dependencies. (C) dependency validation. (D) final $TDG$.} 
\label{fig:example-recovery} 
\end{figure}

%% file: example-independent.tex
\begin{figure}[t]
\centering

\includegraphics[trim={0cm 10cm 0cm 0cm},clip,width=\linewidth]
{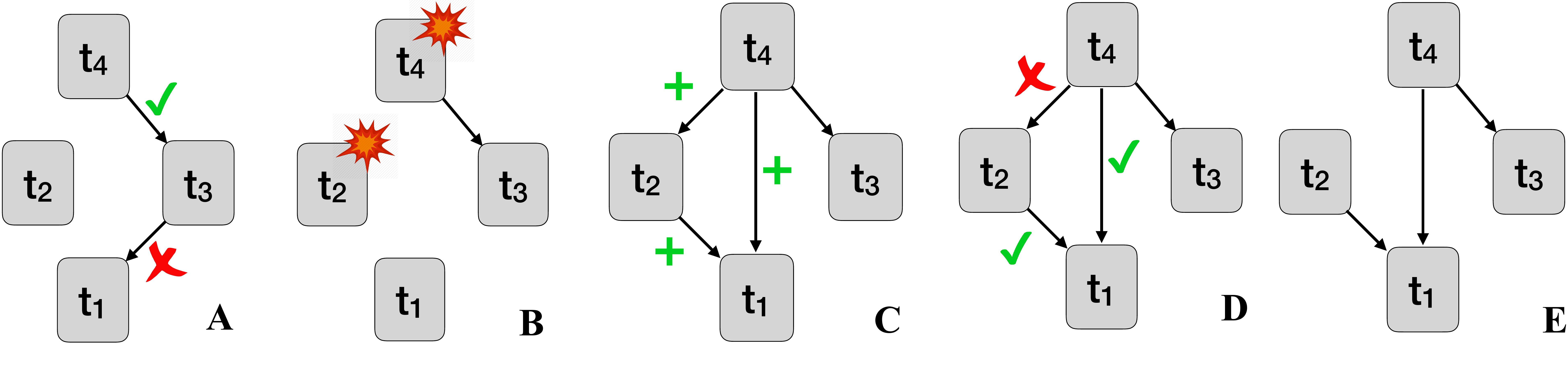}

\caption{Disconnected dependency recovery. (A) dependencies are validated; $t_3 \rightarrow t_1$ shadows the missing dependency $t_4 \rightarrow t_1$. (B) $t_4$ and $t_2$ fail because of missing dependencies. (C) recovery procedure adds candidate dependencies. (D) dependencies are validated. (E) final $TDG$.} 
\label{fig:example-independent} 
\end{figure}

%% file: algo-post-validation-recovery.tex
\begin{algorithm}[h!]
	
	\DontPrintSemicolon
	\scriptsize
	
	\SetKwInOut{Input}{Input}
	\SetKwInOut{Output}{Output}
	\Input{$T_o$: test suite in its original order $o$ \\ $TDG$: test dependency graph}
	\Output{$TDG$: updated test dependency graph with missing dependencies recovered}
	
	\textit{expResults} $\gets$ \textsc{executeTestSuite}($T_o$) \; 
	\HiLi{/* Get isolated nodes and nodes with no outgoing edges. */} \;
	\textit{disconnectedTests} $\gets$ \textsc{getDisconnectedTests}($TDG$) \; 
	\ForEach{\textit{disconnectedTest} in \textit{disconnectedTests} } {
		\textit{execResults} $\gets$ \textsc{executeTestInIsolation}(\textit{disconnectedTest}) \;
		\textit{failedTest} $\gets$ \textsc{getFailedTest}(\textit{expResults}, \textit{execResults})\;
		\uIf{failedTest $\neq$ null}{
			$TDG$ $\gets$ \textsc{connectWithPrecedingTests}(\textit{failedTest}, $TDG$, $T_o$)
		} \ElseIf{\textsc{isNotIsolated}(\textit{disconnectedTest})} {
				\HiLi{/* Out-degree = 0; in-degree $>$ 0. */} \;
				\textit{schedules} $\gets$ \textsc{computeSchedules}(\textit{disconnectedTest}, $TDG$)\;
				\ForEach{schedule $\in$ schedules} {
					\textit{execResults} $\gets$ \textsc{exec}(schedule)\;
					\textit{failedTest} $\gets$ \textsc{getFailedTest}(\textit{expResults}, \textit{execResults})\;
					\If{failedTest $\neq$ null}{
						$TDG$ $\gets$ \textsc{connectWithPrecedingTests}(\textit{failedTest},$TDG$, $T_o$)
					}
				}
		}
	}

	\caption{Disconnected dependency recovery algorithm}
	\label{algorithm:post-validation-recovery}
\end{algorithm}

%% file: 4-evaluation.tex

\section{Empirical Evaluation}\label{sec:evaluation}


We consider the following research questions: 

\noindent
\textbf{RQ\textsubscript{1} (effectiveness):}
How effective is \tool at filtering false dependencies without missing  dependencies to be recovered?

\noindent
\textbf{RQ\textsubscript{2} (performance):}
What is the overhead of running \tool?
What is the runtime saving achieved by \tool with respect to validating \baseline test dependency graphs?

\noindent
\textbf{RQ\textsubscript{3} (parallel test execution):}
What is the execution time speed-up of the test suites parallelized from the test dependency graphs computed by \tool?


\subsection{Subject Systems}

We selected six open-source web applications used in previous web testing research, for which Selenium test suites are available~\cite{WCRE}. 
\autoref{table:subjectSystems} lists our subject systems, including their names, version, size in terms of lines of code,  number of test cases, and the total number of lines of test code. We use \texttt{cloc}~\cite{cloc} to count lines of code. During our experiments we used the original execution order of each test suite, as specified by the developers.


%
%
\input{subjects-table}

\input{table-RQ1-RQ2-RQ3}

\subsection{Procedure and Metrics}

\subsubsection{Procedure}

We manually fixed any flakiness of the test cases of the subject test suites by adding delays where appropriate and we executed each test suite 30~times to ensure that identical outcomes are obtained across all executions.

To form a \textit{baseline} for comparison, we applied dependency validation to the dependency graph obtained from
the original order of each test suite (\autoref{sec:approach-conservative}).

For each  test suite, we ran different configurations of \tool, by combining each admissible combination of graph extraction and filtering technique.
The first evaluated configuration is \textit{String Analysis} (SA), in which the dependency graph is obtained through sub-use chain extraction (\autoref{sec:approach-sa}) and filtered from the dependency-free values (\autoref{sec:approach-sa-filtering}).
Then, we evaluated three configurations in which we applied the three proposed NLP filters (NLP-Verb, NLP-Dobj, NLP-Noun) both to the graph from the original order
as well as to the graph obtained with SA.  

Finally, given the validated dependency graph obtained in each configuration, we generated all possible test \textit{schedules} that respect the test dependencies, and we executed them sequentially. 




\subsubsection{Metrics}


To assess \textit{effectiveness} (RQ\textsubscript{1}), for each configuration we measured the number of \textit{false dependencies} removed by \tool as well as the number of manifest dependencies that are \textit{missing} and need to be \textit{recovered}.
The number of false dependencies is obtained by subtracting the number of manifest dependencies retrieved at the end of the recovery step from the total number of dependencies in the initial graph.



We evaluated \textit{performance} (RQ\textsubscript{2}) by comparing the execution time (in minutes) of each configuration of \tool with respect to the baseline approach.



Concerning \textit{parallelization} (RQ\textsubscript{3}), we measured the speed-up factor of the parallelizable test suites with respect to the original test suite running time.
We considered two speed-up scenarios. (1)~\textit{average case}, in which we measured the ratio between the original test suite running time and the average running time of the parallelizable test suites, and (2)~\textit{worst case}, in which we measured the speed-up ratio between the original test suite running time and the parallelizable test suite having the highest runtime.

\subsection{Results}

\head{Effectiveness (RQ\textsubscript{1})}
For each configuration of \tool, \autoref{table-RQ1-RQ2-RQ3} (Effectiveness) shows
the number of \textit{extracted} dependencies starting from the initial test suite (\autoref{fig:approach}, step \ding{182}) and the number of \textit{filtered} dependencies (\autoref{fig:approach}, step \ding{183}). It also reports information about the validation and recovery steps, specifically the number of \textit{false} dependencies detected, the number of dependencies \textit{recovered} and those recovered from the disconnected components. The final number (Column~8) shows the number of dependencies in the final $TDG$s, all of which are \textit{manifest} \praw dependencies.

Across all apps, the baseline approach validated on average 536 dependencies, of which 504 were deemed as false, and 32 as manifest. 
The most conservative among \tool's configurations is NLP-Verb (Original Order), which validated overall 416 dependencies, of which 384 were false (24\% less than the baseline) and detecting 32 manifest dependencies without filtering/recovering any. 
The least conservative configuration of \tool is NLP-Noun (String Analysis) which retained only 143 dependencies on average from the initial graphs, of which 110 were detected as false, five dependencies had to be recovered, leading to the final number of 33 manifest dependencies. Overall, the number of missing dependencies due to filtering and recovered in steps \ding{184} \ding{185} is very low (1\% of the initial number of dependencies). 

\tool does not ensure having minimal test dependency graphs. Therefore, the number of manifest dependencies retrieved by each configurations is slightly different, between 32 and 34 (Column \textit{Total \praw}). However, these differences do not affect the executability of the schedules that respect the dependencies (see results for RQ\textsubscript{3}). 
%


\input{figure-pareto}

\autoref{fig:pareto} shows the Pareto front plotting the ratio between false and missing dependencies, for each configuration. Each point represents the average $\langle$\textit{missing, false}$\rangle$ values across all subjects, normalized over the respective maximum values.
This essentially shows the tradeoff between the false dependencies remaining after the filtering step and the missing dependencies to be recovered. 

From the analysis of the Pareto front, we can see that the non-dominated configurations are those based on  NLP-Verb, NLP-Noun and String Analysis (SA). The baseline approach (Baseline) has the highest number of false dependencies (536 on average) and no missing dependencies. On the contrary, SA filters many dependencies (393 on average) but has the highest number of missing/recovered dependencies (11 on average). Interestingly, NLP-Verb (Original Order) does not miss any manifest dependency but has more false dependencies compared to the other NLP-based configurations. 
Configurations NLP-Dobj (both SA and Original Order) and NLP-Noun (both SA and Original Order) are comparable regarding the number of false dependencies remaining after  filtering.  However, NLP-Noun (SA and Original Order) needs to recover substantially less manifest dependencies. 
Indeed, NLP-Noun dominates NLP-Dobj, while both NLP-Noun (SA and Original Order) configurations are on the non-dominated front, being both optimally placed in the lower-left quadrant of the Pareto plot.


\head{Performance (RQ\textsubscript{2})}
\autoref{table-RQ1-RQ2-RQ3} (Performance) reports the average runtime, in minutes, of each step of \tool across all configurations. 
The most expensive step of \tool is  validation, especially for what concerns validating the connected part of the graph (Column~12), whereas dependency graph extraction and filtering (Columns~10 and~11) have negligible costs (under one minute on average in all cases). 
The cost of disconnected components recovery (Column~13) is generally low, ranging from nearly one minute for NLP-Verb to maximum nine minutes for NLP-Dobj (3.8 minutes on average).

The slowest configuration of \tool is NLP-Verb (Original Order) which is  27\% faster on average (almost 2~hours less) than the baseline approach. The fastest configuration of \tool is NLP-Noun (SA) which is 72\% faster on average (5~hours less) than the baseline approach. This result confirms the Pareto front analysis,  showing that  NPL-Noun (SA)  is the most effective configuration of \tool.

The table reports also the percentage decrease of each configuration with respect to the baseline, which took approximately 425 minutes on average ($\approx$7~hours).
Overall, all SA- or NLP-based configurations of \tool are significantly faster.

\head{Parallel Test Execution (RQ\textsubscript{3})} 
Column~15 (\textit{schedules}) reports the average number of test schedules obtained from the final $TDG$s. Isolated nodes in the dependency graphs are counted as (single-test) schedules.
Columns~16 and~17 report the relative speed-up of the parallelizable test suites considering the longest test execution schedule (worst-case) and the average case. 

First, in our experiments, no test failures occurred in any of the 
parallelizable test suite produced by any configuration of \tool. Essentially, this testifies that the dependency validation and recovery algorithm does not miss any manifest dependency. 

Overall, all techniques achieve similar speed-up scores, around 2$\times$ in the worst case and 7$\times$ on average. This is expected since the final TDGs are similar across configurations (see total number of manifest dependencies in \autoref{table-RQ1-RQ2-RQ3}). However, results differ across applications. \autoref{table:RQ3} presents the results for the NLP-Noun (SA) configuration. Column \textit{runtime original} reports the execution time of the original test suite in seconds.
\textit{Collabtive} has the slowest test suite (almost 5 minutes) whereas \textit{Addressbook} has the fastest (40 seconds). 
The highest speed-up in the worst-case occurs for \textit{Claroline}, where the 
longest test execution schedule test suite is 2.7$\times$ faster. 
\textit{MantisBT} exhibits the highest speed-up in the average case (12.3$\times$), but the lowest speed-up in the worst-case (1.4$\times$) due to a single slow-executing schedule (133 s) with respect to the average runtime (15 s). 
The lowest speed-up in the average case occurs for \textit{MRBS}, but it remains still high (4.2$\times$).

\input{table-RQ3}

%% file: subjects-table.tex
\begin{table}[h!]
\setlength{\tabcolsep}{5pt}
\renewcommand{\arraystretch}{0.9}
\centering
\caption{Subject systems and their test suites}
\begin{tabular}{lll@{\hskip 2em}c@{\hskip 2em}c}
\toprule

& \multicolumn{2}{c}{\sc Web App} 
& \multicolumn{2}{c}{\sc Test Suites} \\

\cmidrule(r){2-3} \cmidrule(r){4-5}

& Version
& LOC
& \#
& LOC (Avg/Tot) \\

\midrule

Claroline & \texttt{\small 1.11.10} & 352,537 & 40 & 46/1,822         \\
AddressBook & \texttt{\small 8.0.0} & 16,298 & 27 & 49/1,325          \\
PPMA & \texttt{\small 0.6.0} & 575,976 & 23 & 54/1,232          \\
Collabtive & \texttt{\small 3.1} & 264,642 & 40 & 48/1,935          \\
MRBS & \texttt{\small 1.4.9} & 34,486 & 22 &  51/1,114        \\
MantisBT & \texttt{\small 1.1.8} & 141,607 & 41 & 43/1,748          \\

\midrule
Total & & 866,995 & 196 & 47/9,176 \\

\bottomrule
\end{tabular}
\label{table:subjectSystems}
\end{table}

%
%
%
%
%

%% file: table-RQ1-RQ2-RQ3.tex
\begin{table*}[]

\begin{threeparttable}[b]

\caption{Effectiveness (RQ1), Performance (RQ2) and Parallelization (RQ3) average results across all subject test suites.}
\label{table-RQ1-RQ2-RQ3}
\setlength{\tabcolsep}{3.6pt}
\renewcommand{\arraystretch}{0.9}

\begin{tabular}{l@{\hskip 1em}llllllll@{\hskip 2em}llllll@{\hskip 2em}l@{\hskip 1.5em}cc}

\toprule 

& \multicolumn{8}{c}{\textsc{Effectiveness}}  
& \multicolumn{6}{c}{\textsc{Performance}}
& \multicolumn{3}{c}{\sc Parallelization} \\

\cmidrule(r){2-9} 
\cmidrule(r){10-15}
\cmidrule(r){16-18}

&           &          &             &       & \multicolumn{4}{c}{\bf Manifest Deps.}                                      &  &  & \multicolumn{2}{c}{\bf Validation}  &  & &  & \multicolumn{2}{c}{\bf Speed-up (\%)} \\


\cmidrule(r){6-9} 
\cmidrule(r){12-13} 
\cmidrule(r){17-18} 

& \rot{Extracted}
& \rot{Filtered}
& \rot{To Validate}
& \rot{False}
& \rot{Validated}
& \rot{Recovered}
& \rot{Recovered (Disc.)}
& \rot{Total \praw}
 
& \rot{Extraction} 
& \rot{Filtering} 
& \rot{Val. and Recovery}
& \rot{Recovery (Disc.)}
& \rot{Total}
& \rot{Saving (\%)} 

& \rot{Schedules (\#)} 
& \rot{Worst-case} 
& \rot{Average} \\

\midrule

Baseline (Original Order)                   & 536       & -        & 536         & 504   & 32        & 0                & 0              & 32    & 0.00\tnote{$\dagger$}       & -         & 424.7\tnote{$\ast$}         & -             & 424.70 & -           & 30          & 2.2$\times$        & 7.1$\times$      \\
String Analysis            & 494       & 393      & 101         & 69    & 21        & 10               & 1              & 32    & 0.10       & 0.00      & 162.02          & 5.21          & 167.33 & 61\%        & 30          & 2.4$\times$        & 7.1$\times$     \\ [0.3em]
NLP-Verb (Original Order)        & 535       & 119      & 416         & 384   & 32        & 0                & 0              & 32    & 0.00\tnote{$\dagger$}       & 0.04      & 307.20          & 1.27          & 308.51 & 27\%        & 30          & 2.2$\times$        & 7.1$\times$      \\
NLP-Verb (String Analysis) & 494       & 113      & 381         & 348   & 32        & 1                & 0              & 33    & 0.09       & 0.04      & 281.10          & 1.28          & 282.50 & 33\%        & 30          & 2.2$\times$        & 7.1$\times$      \\[0.3em]
NLP-Dobj (Original Order)        & 536       & 362      & 174         & 140   & 27        & 6                & 1              & 34    & 0.00\tnote{$\dagger$}       & 0.24      & 134.90          & 3.46          & 138.60 & 67\%        & 29          & 2.1$\times$        & 6.7$\times$      \\
NLP-Dobj (String Analysis) & 494       & 343      & 151         & 117   & 27        & 5                & 2              & 34    & 0.09       & 0.23      & 129.20          & 9.10          & 138.62 & 67\%        & 29          & 2.1$\times$        & 6.7$\times$      \\[0.3em]
NLP-Noun (Original Order)        & 536       & 364      & 172         & 140   & 28        & 3                & 1              & 32    & 0.00\tnote{$\dagger$}       & 0.05      & 123.30          & 2.52          & 125.87 & 70\%        & 29          & 2.1$\times$        & 6.7$\times$      \\
NLP-Noun (String Analysis) & 494       & 351      & 143         & 110   & 28        & 4                & 1              & 33    & 0.08       & 0.04      & 116.10          & 4.08          & 120.30 & 72\%        & 29          & 2.1$\times$        & 6.8$\times$     \\

\bottomrule

\end{tabular}

\begin{tablenotes}[para]
\item[$\ast$] \textit{only validation, no within-recovery.}
\item[$\dagger$] \textit{execution time $<$ 0.01 minutes (0.6 seconds).}
\end{tablenotes}
\end{threeparttable}

\end{table*}

%% file: figure-pareto.tex
\begin{figure}[t]
\centering

\includegraphics[trim={5.8cm 2cm 8cm 2cm},clip,width=\linewidth]
{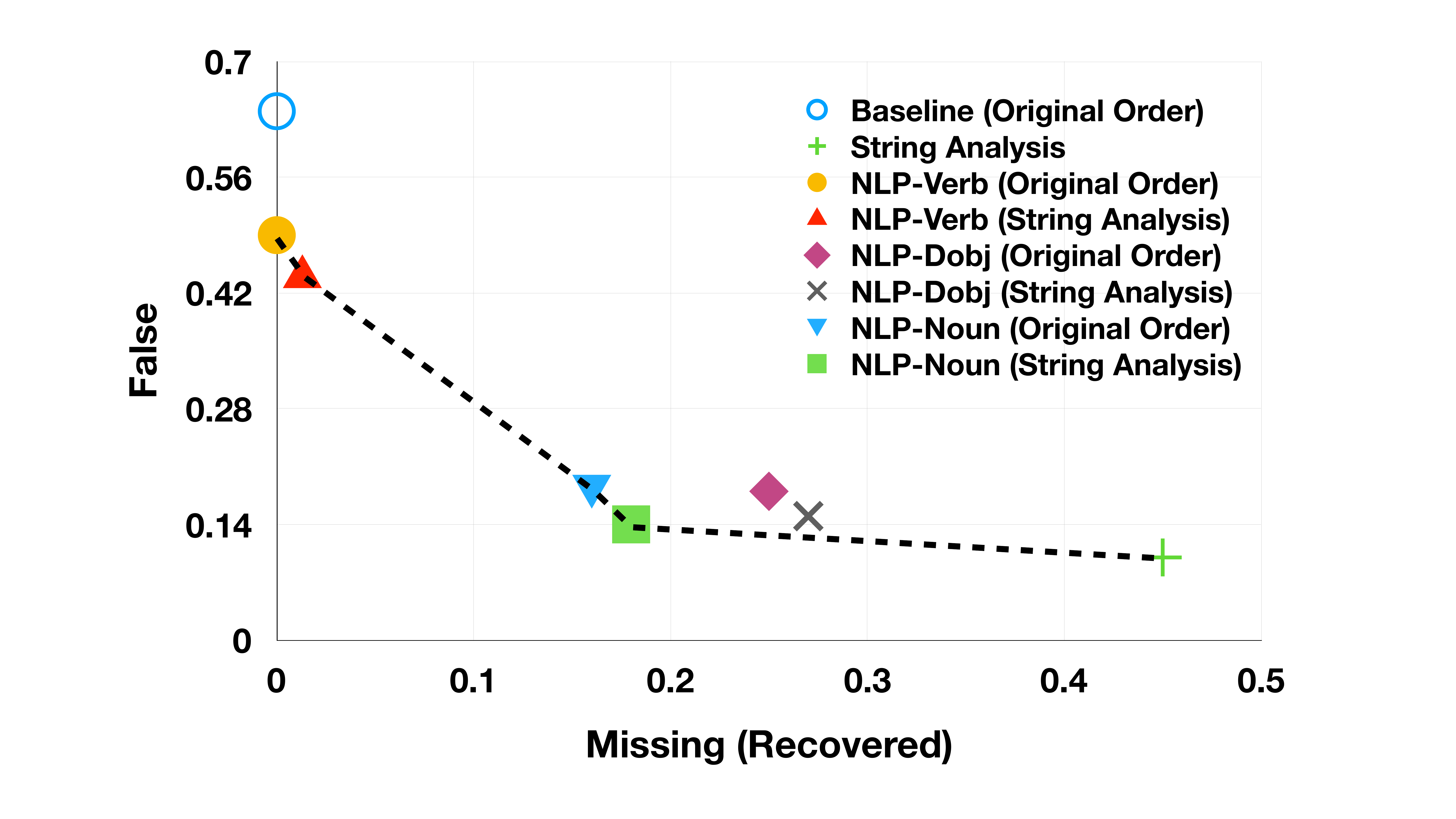}
\caption{Pareto front} 
\label{fig:pareto} 
\end{figure}

%% file: table-RQ3.tex
\begin{table}[t]

\begin{threeparttable}[b]

\caption{Parallelization results for NLP-Noun (SA)}
\label{table:RQ3}
\setlength{\tabcolsep}{5.6pt}
\renewcommand{\arraystretch}{0.9}

\begin{tabular}{lllllll}

\toprule



&  &  & \multicolumn{2}{c}{\bf Worst-case} & \multicolumn{2}{c}{\bf Average} \\

\cmidrule(r){4-5}
\cmidrule(r){6-7}  

& \rot{runtime original} 
& \rot{schedules (\#)} 
& \rot{runtime} 
& \rot{speed-up (\%)} 
& \rot{runtime} 
& \rot{speed-up (\%)} \\


\midrule

Claroline & 75.1 & 36 & 29.0 & 2.7$\times$ & 9.4 & 8.3$\times$ \\
Addressbook & 40.2 & 24 & 20.1 & 2.0$\times$ & 8.8 & 4.5$\times$ \\
PPMA & 51.7 & 22 & 21.1 & 2.4$\times$ & 8.9 & 5.8$\times$ \\
Collabtive & 297.7 & 37 & 133.1 & 2.3$\times$ & 53.2 & 5.7$\times$ \\
MRBS & 56.9 & 20 & 28.7 & 2.0$\times$ & 13.8 & 4.2$\times$ \\
MantisBT & 184.5 & 37 & 133.8 & 1.4$\times$ & 15.1 & 12.3$\times$ \\

\midrule

Total & 706.1 & 176 & 365.9 & 2.1$\times$\tnote{*} & 109.2 & 6.9$\times$\tnote{*} \\

\bottomrule
\end{tabular}

\begin{tablenotes}
\item[*] \textit{average}
\end{tablenotes}
\end{threeparttable}

\end{table}

%% file: 5-discussion.tex
\section{Discussion}\label{sec:discussion} 


\head{Automation and Effectiveness} 
Our results confirm that (1)~E2E web tests entail test dependencies, (2)~such dependencies can be identified by considering \praw connections between test cases, and (3) 
\tool can successfully detect all \praw test dependencies necessary for independent test case execution. All proposed filtering techniques proved both very fast and effective at reducing the size of the initial graph, without filtering many manifest dependencies. 

\head{Performance and Overhead}
All configurations of \tool achieve substantial improvements with respect to validating the graph extracted from the original ordering, whose validation cost is quadratic on the number of test cases. This means that every time a new test case is added to a test suite containing $n$ test cases, such graph requires that $n$ more dependencies are validated. 
Moreover, despite the computational cost of NLP processing, the dependency graph extraction and filtering steps exhibit negligible costs (less than a minute over all test suites) compared to the validation time.



\head{Relation to Test Optimization Techniques}
\tool can be used by test engineers to detect \praw-like dependencies in E2E web test cases. 
The validated dependency graph produced by \tool can be used as input for devising novel  test optimization and regression testing techniques in the web domain. 

For instance, test prioritization and test minimization can be formulated as constrained optimization problems, in which dependencies by \tool play the role of constraints that can be addressed using SMT solvers~\cite{z3} or search-based heuristics~\cite{sbse}. 
Test selection can benefit from the $TDG$s produced by \tool to identify which test cases are required by modification-traversing test cases~\cite{Rothermel:1997:SER:248233.248262,536955,STVR197}.

In this work, we studied an application to test parallelization as a proxy to assess the correctness of the $TDG$s produced by \tool. Specifically, we applied dependency-aware graph traversal to retrieve all potential parallel test suites, and measured the runtime speed-up with respect to the initial test suites. 

\head{Test Smells}
Our test dependency graph can also be utilized for other test analysis activities such as detecting \textit{poorly designed} tests (i.e., test smells~\cite{van2001refactoring}), or detecting obsolete tests. 

For instance, the test case \texttt{checkEntryTagsRemoved} of  \textit{PPMA} executes after \mbox{\texttt{addEntryTags}} and \texttt{removeEntryTags} tests. By analyzing the $TDG$ produced by \tool for this test suite, we noticed that \texttt{checkEntryTagsRemoved} executes properly also when no tags have been created yet (i.e., \texttt{checkEntryTagsRemoved} is \textit{isolated} in the $TDG$). 
This means that the test \texttt{checkEntryTagsRemoved} is obsolete because subsumed by the previous \texttt{removeEntryTags} test. Therefore, 
it can be safely removed with no impact on the functional coverage or assertion coverage of the test suite. 

\head{Test Suite Evolution} 
During software evolution, 
if new test cases are added to the test suite, dependencies are  added to the previously validated $TDG$ through new dependency extraction and filtering. 
Validation and recovery must be carried out on the entire new $TDG$. However, this analysis is expected to be faster than the first initial one, given that the validated $TDG$ we are adding dependencies to contains only manifest dependencies. 

\head{Limitations}
\tool depends on the information available in the test source code, used to identify potential test dependencies. 
As such, the effectiveness of our NLP-based filtering may be undermined if test case names are not descriptive, as in the case of many automatically generated test suites (e.g., \code{test1}, \code{test2}). 
In such cases, testers can rely on the string analysis configuration of \tool (SA), which also proved effective in our study. 
Second, our tool does not provide information about the \textit{root cause} of the dependencies, i.e., what part of the program state is polluted by which test.
%
Lastly, \tool does not support the analysis of flaky test suites.

\head{Threats to Validity}
Using a limited number of test suites in our evaluation poses an \textit{external validity} threat. 
Although more subject test suites are needed to fully assess the generalizability of our results, we have chosen six subject apps used in previous web testing research, pertaining to different domains, for which test suites were developed by a human web tester. 

Threats to \textit{internal validity} come from confounding factors of our experiments, such as test flakiness. 
To cope with possible flakiness of the test cases, we manually fixed any flaky test by adding delays where appropriate and we ran each test suite 30 times to ensure having identical results on all executions.

%% file: 6-related.tex
\section{Related Work}\label{sec:relwork}

\head{Test Dependency Detection}
Different techniques have been proposed recently to detect dependencies in unit tests, none of which focuses on web tests.
Zhang et al.~\cite{Zhang:2014:ERT:2610384.2610404} developed DTDetector, which detects manifest dependencies in JUnit tests with a dependency-aware $k$-bounded algorithm. They showed that a small value for $k$ (e.g., $k=1$ and $k=2$) finds most realistic dependent tests.
Poldet~\cite{Gyori:2015:RTD:2771783.2771793} retrieves tests that pollute the shared state of a Java application (i.e., heap or filesystem), and provides ways to inspect such polluted states, e.g., access path through the heap that leads to the modified value, or the name of the file that was modified.
The tool VMVM~\cite{Bell:virtual} uses test virtualization to isolate unit tests of a JUnit suite, by resetting the static state of the application to its default before each test execution. Our tool \tool, differently from the previous works, targets the web domain by focusing on  dependencies due to read-after-write operations performed on persistent data. 
ElectricTest~\cite{Bell:2015:EDD:2786805.2786823} utilizes dynamic data-flow analysis to identify all conflicting write and read operations over static Java objects. On the contrary, in E2E web tests, the semantics of read and write operations is implicit and  mediated by multiple layers of indirection such as client-side DOM, server-side application state, database entries, and remote service calls. \tool leverages heuristics based on sub-use-chain relations and NLP to discover potential test dependencies without the need for complex data-flow analysis. 

Pradet~\cite{pradet} detects test dependencies in Java unit tests focusing on manifest dependencies that can be traced back to data dependencies. \tool adopts a similar approach as Pradet to validate the test dependencies, i.e., validating a single dependency at a time by inverting the dependency and linearizing the graph.  

The main differences between Pradet and our approach are (1)~the capability of \tool to handle incomplete dependency graphs; Pradet makes the assumption that the initial dependency graph contains all manifest dependencies computed through static analysis of all (read/write) accesses to global Java variables. Due to this assumption, (2)~Pradet only removes false dependencies and it does not recover any missing dependencies. \tool, on the other hand, features a novel recovery algorithm to detect and validate all potentially missing manifest dependencies, which makes it suitable for web E2E test suites, in which applying thorough data-flow analysis is neither feasible nor straightforward, due to the heterogeneity of technologies and languages used in modern web applications. Lastly, (3)~\tool introduces novel extraction and filtering heuristics, which extend the applicability of Pradet beyond read/write operations performed on static fields of Java classes. 

%% file: 7-conclusion.tex
\section{Conclusions and Future Work}\label{sec:conclusions}

In this paper, we proposed a novel test dependency technique for E2E web test cases implemented in a tool called \tool. 
We used \tool for detecting test dependencies in six web test suites. 
Thanks to an effective combination of novel string analysis and NLP, \tool achieves an optimal trade off between false dependencies to be removed 
and missing dependencies to be recovered. 
Moreover, \tool detected the final set of manifest test dependencies on average 57\% faster than the baseline approach. The resulting test dependency graph supports parallelization of the tests, with a speed-up factor of up to 7$\times$.
In our future work, we plan to apply NLP on the DOM and on page object-based test suites~\cite{mf-po}, as well as to devise ways to produce a \textit{minimal} $TDG$, i.e., a $TDG$ having the minimum number of dependencies. We also intend to run \tool on more subject web test suites and to extend it to support mobile test suites.